# Optimal Variable Speed Limit Control Strategy on Freeway Segments under Fog Conditions

**Ben Zhai**
Key Laboratory of Road and Traffic Engineering of Ministry of Education, Tongji University
No. 4800 Cao'an Road, Shanghai, 201804, China
Email: zhaiben@tongji.edu.cn

**Yanli Wang, Corresponding Author**
Key Laboratory of Road and Traffic Engineering of Ministry of Education, Tongji University
No. 4800 Cao'an Road, Shanghai, 201804, China
Email: wangyanli@tongji.edu.cn

**Wenxuan Wang**
Key Laboratory of Road and Traffic Engineering of Ministry of Education, Tongji University
No. 4800 Cao'an Road, Shanghai, 201804, China
Email: wangwenxuan@tongji.edu.cn

**Bing Wu**
Key Laboratory of Road and Traffic Engineering of Ministry of Education, Tongji University
No. 4800 Cao'an Road, Shanghai, 201804, China
Email: wubing@tongji.edu.cn

Word Count: 5,671 words + 7 table (250 words per table) = 7,421 words

*Submitted [Submission Date]*

**ABSTRACT**

Fog is a critical external factor that threatens traffic safety on freeways. Variable speed limit (VSL) control can effectively harmonize vehicle speed and improve safety. However, most existing weather-related VSL controllers are limited to adapt to the dynamic traffic environment. This study developed optimal VSL control strategy under fog conditions with fully consideration of factors that affect traffic safety risks. The crash risk under fog conditions was estimated using a crash risk prediction model based on Bayesian logistic regression. The traffic flow with VSL control was simulated by a modified cell transmission model (MCTM). The optimal factors of VSL control were obtained by solving an optimization problem that coordinated safety and mobility with the help of the genetic algorithm. An example of I-405 in California, USA was designed to simulate and evaluate the effects of the proposed VSL control strategy. The optimal VSL control factors under fog conditions were compared with sunny conditions, and different placements of VSL signs were evaluated. Results showed that the optimal VSL control strategy under fog conditions changed the speed limit more cautiously. The VSL control under fog conditions in this study effectively reduced crash risks without significantly increasing travel time, which is up to 37.15% reduction of risks and only 0.48% increase of total travel time. The proposed VSL control strategy is expected to be of great use in the development of VSL systems to enhance freeway safety under fog conditions.

*Ben Zhai, Yanli Wang, Wenxuan Wang and Bing Wu*

**INTRODUCTION**

The visibility reduction due to fog is one of the most influential factors in fatal crashes on freeways. In the United States, there are about 300-400 fog related fatal crashes happening every year (*1*). Crashes under fog conditions, involving large number of vehicles, tend to be more severe (*2*). Therefore, it is necessary to devote efforts to carry out effective measures to improve the traffic safety under fog conditions on freeway.

Variable speed limit (VSL) control has been increasingly applied as an active traffic management strategy to harmonize vehicle speed and improve safety on freeways under adverse weather conditions (*3-5*). Indicated by VSL signs on the roadside, drivers are advised to adjust their speed in advance and take the appropriate speed to pass through the controlled road sections. VSL control can lower the speed variance among vehicles and the speed difference between upstream and downstream, so as to improve traffic safety (*6; 7*). In addition, the reduction in speed differences synchronizes the behaviors of drivers, and the lane-changing behavior is discouraged (*8*). As a result, the reduction of collision probability can be achieved by VSL control (*9-11*).

Previous studies concluded that VSL control significantly reduced traffic safety risk under adverse weather conditions. Elvik et al. suggested that VSL control reduced accidents by 13%, and weather-based VSL control could reduce accident probability by 2% (*12*). Saha et al. found that VSL control had a certain impact on the probability of accidents, with a reduction of 0.78 accidents on every 100 miles of VSL corridors per week (*13*). Gonzales et al. evaluated the effects of the I-77 fog VSL system in Fancy Gap, Virginia on driving speed (*14*). The results indicated that VSL control significantly reduced the mean speeds, and drivers drove closer to the safe speed based on available visibility. The central idea of most practical weather-related VSL control systems was to adjust the speed limit to a prespecified value if the current weather conditions were worse than ideal. However, the speed limits of most control strategies were usually pre-set fixed values for several levels of weather conditions (*15-17*), and were usually determined according to practical experiences (*18*), which was limited to adapt to the dynamic and complex traffic environment.

Some studies have conducted dynamic methods to develop the VSL control strategy under fog conditions based on the influence of visibility reduction on driving behavior. FHWA proposed a dynamic algorithm to calculate the real-time maximum safe speed limit in adverse weathers with the consideration of weather-related factors such as the road adhesion and sight distance (*19*). Choi et al. reported a novel VSL operations strategy based on the prediction of weather and traffic conditions (*20*). Li et al. introduced a strategy of VSL based on modified car-following models to reduce secondary collisions in adverse weather (*18*). This strategy adjusted speed limit according to real-time traffic and weather conditions. In fact, traffic crash risks are affected by various factors including traffic flow, traffic management measures, road environment, and weather factors (*21-23*). Although these micro-based control strategies directly reflected the impact of reduced visibility on sight distance and stopping distance, the interactions among traffic flow status, road geometric characteristics and safety risks were not fully considered, which leads to an inaccurate estimation of crash risk under fog conditions, thereby weakening the benefits of VSL control.

The primary objective of this study is to develop an optimal VSL control strategy under fog conditions with consideration of factors that affect traffic safety risks including visibility, traffic flow state, and road geometric characteristics. The framework of determining optimal VSL control strategy under fog conditons considered safety and mobility. A simulation experiment of I-405 in California, US was designed to evaluate the effects of the proposed VSL control strategies. The effects of different placements of VSL signs ware also evaluated. This study can help transportation managers to design VSL control systems to achieve safety improvement under fog conditions.

**METHODOLOGY**

The architecture of optimal VSL control strategy under fog conditions is shown as **Figure 1**. The relationship between crash risk and various factors was examined based on a crash risk prediction model under fog conditions to achieve safety evaluation. A modified cell transmission model (MCTM) was established to simulate the evolution of the traffic flow and evaluate the mobility. Then, a model predictive control (MPC) based method of determining the optimal VSL control strategy was developed, and the genetic algorithm (GA) was adopted to searching for the optimal combination of critical factors.

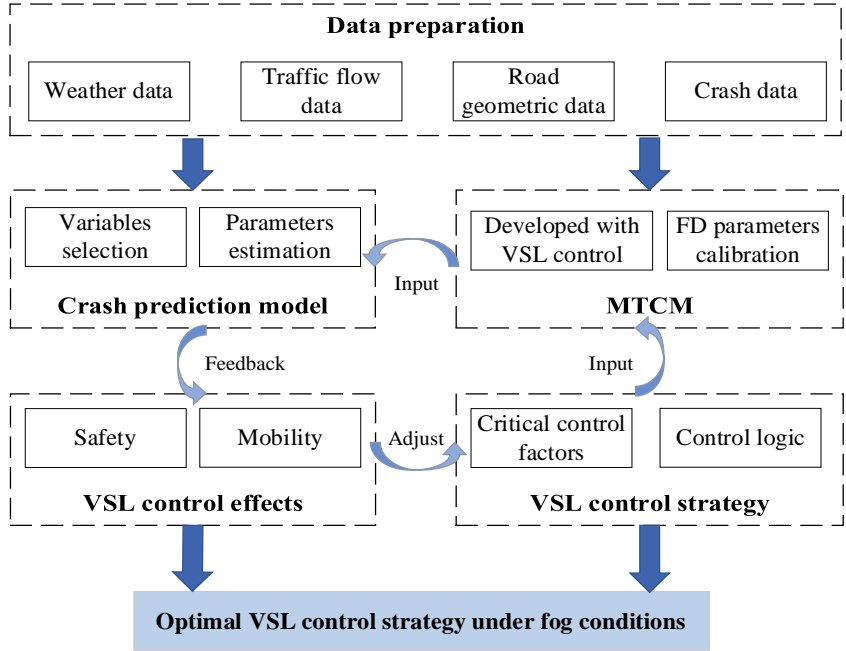

**Figure 1 Structure for determining optimal VSL control strategy**

**Crash Risk Prediction Model under Fog Conditions**
Crash risk can be used to effectively evaluate the safety effects of the VSL control strategy. To this end, a relationship between crash risk and its related factors needs to be established. Previously, many researchers applied risk prediction models based on real-time loop detector data to assess traffic safety risks under specific traffic flow conditions (*24-28*). However, few studies focus on the relationship between traffic flow and crash risk under fog conditions. In this paper, the crash risk prediction model under fog conditions that was introduced in the authors' previous study (*26*) was used to assess the crash risk given real-time traffic flow parameters, road geometric characteristics and weather data. The development of the crash risk prediction model was briefly discussed in this section.

Bayesian logistic regression model was adopted to build the real-time crash risk assessment model. It can be written as:

$$y_i \sim \text{Bernoulli}(p_i) \quad (1)$$

$$p_i = P(y_i = 1) = \frac{e^{\eta_i}}{1 + e^{\eta_i}} \quad (2)$$

$$\eta_i = \beta_0 + \sum_{j=1}^{k} \beta_j x_{ji} \quad (3)$$

where $y_i$ represents the crash indicator for the *i*th observation in the sample; $p_i$ denotes the crash probability for the *i*th observation; $\eta_i$ is the multiple linear combination of explanatory variables; $x_{ji}$ represents the value of variable *j* for sample *i*; $\beta_j$ is the coefficient of variable *j*; and $\beta_0$ is the intercept. The likelihood function can be written as:

$$L(\beta) = \prod_{i=1}^{N} p_i = \prod_{i=1}^{N} \left[ \left( \frac{e^{\eta_i}}{1+e^{\eta_i}} \right)^{y_i} \left( 1 - \frac{e^{\eta_i}}{1+e^{\eta_i}} \right)^{(1-y_i)} \right] \quad (4)$$

Based on the specification of the Bayes theorem, the posterior distribution of parameters can be estimated using the following function:

$$f(\beta_0, \cdots, \beta_j | y) \propto f(y|\beta_0, \cdots, \beta_j) f(\beta_0, \cdots, \beta_j)$$

$$\propto \prod_{i=1}^{N} \left[ \left( \frac{e^{\eta_i}}{1+e^{\eta_i}} \right)^{y_i} \left( 1 - \frac{e^{\eta_i}}{1+e^{\eta_i}} \right)^{(1-y_i)} \right] \prod_{j=0}^{k} \left[ \frac{1}{\sqrt{2\pi}\sigma_j} e^{-\frac{1}{2}\left(\frac{\beta_j - \mu_j}{\sigma_j}\right)^2} \right] \quad (5)$$

In each cycle of the VSL control, the risk value is calculated based on real-time visibility data, traffic flow parameters, and road geometric information. Then, VSL control is determined by the comparison of the risk value in the current state and a prespecified threshold. The risk threshold is set to be 0.2 according to the clustering result with fuzzy C-means algorithm in (*29*).

**Development of Traffic Simulation Model**
Macroscopic traffic simulation models have been widely applied in active and real-time traffic management systems, because they can capture many important traffic phenomena and model a large-scale network with heterogenic elements in an efficient and stable manner. Among macroscopic models carried out in literature, the Cell Transmission Model (CTM) is one of the most widely deployed traffic flow models, which is introduced by Daganzo (*30; 31*). The CTM provides a discretization scheme of the Lighthill-Whitham-Richards (LWR) model. It is a special case of the Godunov's scheme for a triangular fundamental diagram (*32*). The CTM is suitable for real-time application on traffic management on freeway because of its concise formulation and strong robustness (*33*).

*Modified cell transmission model*
In Daganzo's version of CTM, the road is divided into homogeneous cells, and the lengths of the cells are set equal to the distances traveled at free-flow speed in a time interval. The CTM was extended to a density-based version entitled the MCTM by Pan et al. (*34*). The MCTM uses traffic density as state variable and allows unequal cell lengths, which leads to greater flexibility in modeling freeway traffic flow. The MCTM was employed in this study to simulate and evaluate the freeway traffic flow.

**TABLE 1 Variables and Parameters of MCTM**

| Symbol | Description |
| --- | --- |
| $\Delta T$ | Simulation time step |
| $n$ | Number of cells |
| $k$ | Discrete time index |
| $l_i$ | Length of cell $i$ |
| $\rho_i(k)$ | Density of cell $i$ in the $k$th interval |
| $q_i(k)$ | Total flows entering cell $i$ in the $k$th interval |
| $r_i(k)$ | On-ramp flow entering cell $i$ in the $k$th interval |
| $f_i(k)$ | Off-ramp flow leaving cell $i$ in the $k$th interval |
| $Q_M$ | Maximum allowable flow |
| $v_f$ | Free-flow speed |
| $w_c$ | Backward congestion wave speed |
| $\rho_c$ | Critical density |
| $\rho_j$ | Jam density |

The related variables and parameters of MCTM are shown in **TABLE 1**. The MCTM divides a section of freeway into *n* cells and each cell has one on-ramp and one off-ramp, as shown in **Figure 2 (*a*)**. Each cell is characterized by a fundamental diagram (FD), which represents the relationship between traffic flow and density (**Figure 2 (*b*)**). In addition, one requirement of the MCTM is that the cell lengths should be longer than the free-flow travel distance in a time interval, that is, $l_i \geq v_{f,i} \Delta T$.

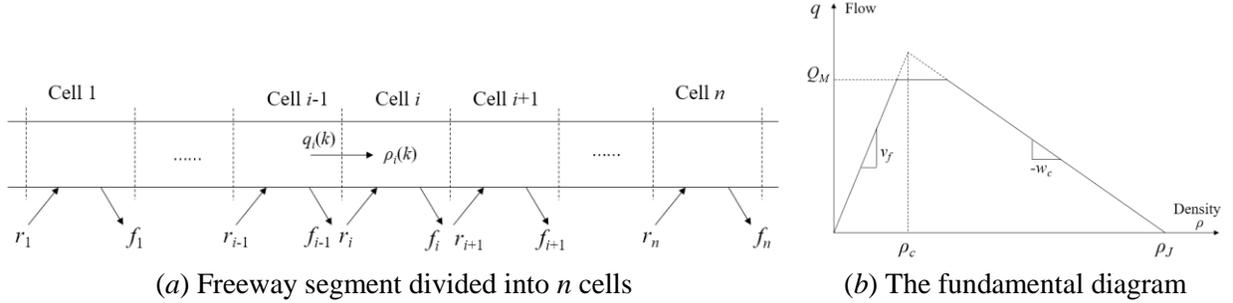

(*a*) Freeway segment divided into *n* cells  (*b*) The fundamental diagram

**Figure 2 Schematic diagram of MCTM**

*Development of MCTM with VSL control*
For the density evolution of cell *i*, it is determined by the density of cell *i* in the previous interval, the flow entering and leaving the cell, as shown in **Equation 6**. $r_i(k)$ and $f_i(k)$ are measured by detectors.

$$\rho_i(k+1) = \rho_i(k) + \frac{\Delta T}{l_i}\left(q_i(k) - q_{i+1}(k) + r_i(k) - f_i(k)\right) \quad (6)$$

$$q_i(k) = \min\{S_{i-1}(k), R_i(k)\} \quad (7)$$

With the control of VSL, the sending function $S_i(k)$ and receiving function $R_i(k)$ are constrained by the reduced speed limit:

$$S_i(k) = \min\{V_{SL,i}(k)\rho_i(k) - f_i(k), Q_{VSL,i}(k) - f_i(k)\} \quad (8)$$

$$R_i(k) = \min\{Q_{VSL,i}(k) - r_i(k), w_{c,i}(\rho_{j,i} - \rho_i(k)) - r_i(k)\} \quad (9)$$

where $V_{SL,i}(k)$ is the speed limit (mph) posted on the VSL sign at time *k*, and $Q_{VSL,i}(k)$ is the maximum flow (veh/h) under current speed limit at time *k*. The speed within cell *i* can be calculated according to the current density and speed limit:

$$v_i(k) = \begin{cases} \min\{v_{f,i}, V_{SL,i}(k-1)\}, & \text{if } \rho_i(k) \leq \rho_{VSL,i}(k) \\ w_{c,i}(\rho_{j,i} - \rho_i(k))/\rho_i(k), & \text{if } \rho_i(k) > \rho_{VSL,i}(k) \end{cases} \quad (10)$$

where $\rho_{VSL,i}(k)$ is the density (veh/mile/ln) associated with the flow $Q_{VSL,i}(k)$ under the speed limit $V_{SL,i}(k)$.

For the two boundary conditions, the MCTM considers the congested conditions at downstream boundaries so as to work with real freeway data (*35*), as shown in **Equation 11 to 14**. Boundary traffic flows and densities are measured and denoted as $q_u^*(k)$, $\rho_u(k)$ and $q_d^*(k)$, $\rho_d(k)$, respectively. Thus, the density of each cell in each time interval is determined once the initial values are given.

$$q_1(k) = \min\{q_u(k), R_1(k)\} \quad (11)$$

$$q_{n+1}(k) = \min\{S_n(k), q_d(k)\} \quad (12)$$

$$q_u(k) = \begin{cases} q_u^*(k), & \text{if } \rho_u(k) \leq \rho_{c,1} \\ \infty, & \text{otherwise} \end{cases} \quad (13)$$

$$q_d(k) = \begin{cases} \infty, & \text{if } \rho_d(k) \leq \rho_{c,n} \\ q_d^*(k), & \text{otherwise} \end{cases} \quad (14)$$

*Calibration of MCTM parameters*
Calibration of the MCTM is to determine the FD of each cell, which represents the relationship between traffic flow and density. In the calibration process, the parameters are regulated until there is a correct relation between the data collected from observation and those reproduced by the model. A two-step calibration method of the MCTM was introduced by this paper. First, the least square method (LSM) was used to estimate the initial value of parameters considering the trapezoidal fundamental diagram (*36*). Second, the initial parameters were put into an optimization problem. Constrained optimization by linear approximation (COBYLA) was used to search for the optimal value of parameters of the FDs. **Figure 3** shows the calibration procedure.

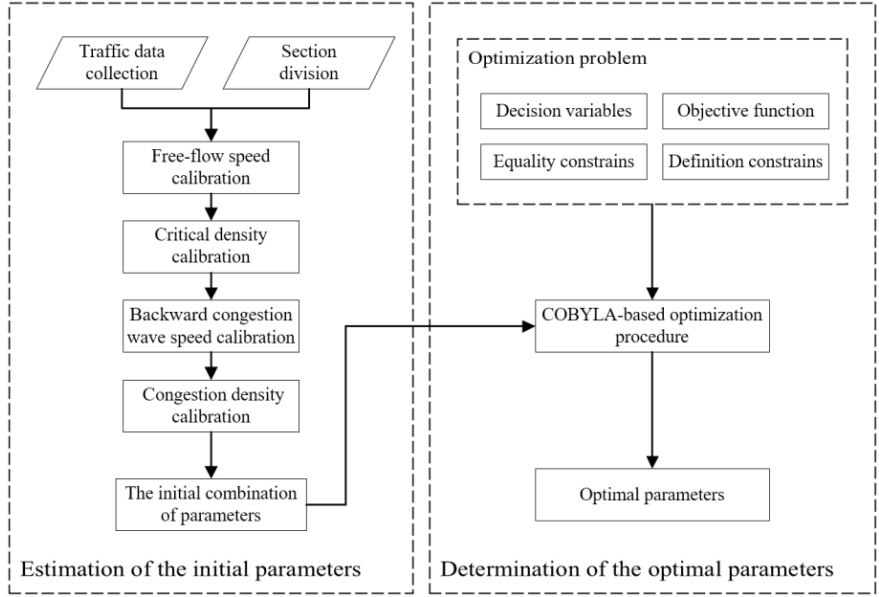

**Figure 3 The calibration procedure of MCTM**

**Optimal VSL Control Strategy**
*Development of VSL control strategy*
It is expected to be a gradual process that the speed limit value displayed on the VSL sign changes from the default value to the target value. If the speed limit changes too frequently, the driver may not have enough time to react or might take dangerous operations such as emergency braking, thereby increasing the risk of traffic accidents. In addition, the excessive difference between the displayed speed limit of adjacent VSL signs may cause the driver to frequently accelerate and decelerate in a short time, resulting in an increase of safety risks (*26*). Therefor, it is necessary to carry out certain constraints from the two dimensions of time and space to guide the driver to pass the controlled road section with a smooth speed change. **Figure 4** shows the control logic of the VSL control strategy proposed in this study. It contains four critical control factors: (1) the start threshold ($R$); (2) the control cycle of VSL ($T$); (3) the speed change step ($\Delta v$); and (4) the maximum speed difference between adjacent signs ($\Delta V_m$).

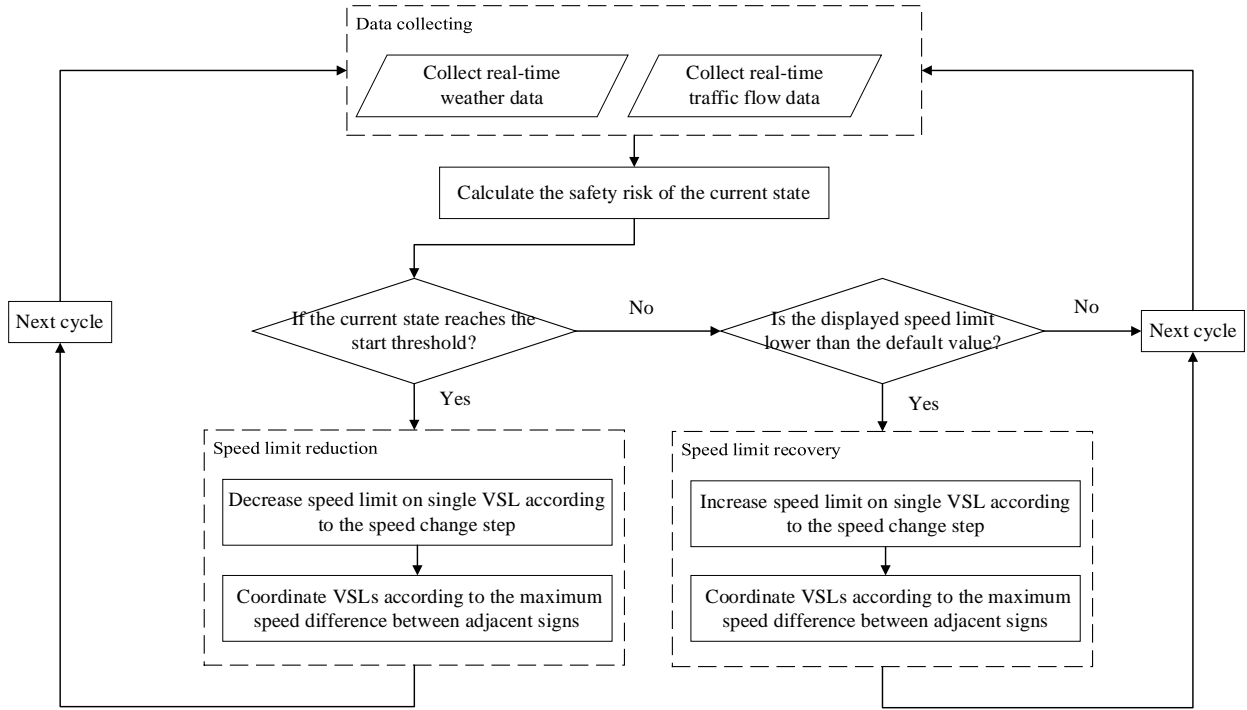

**Figure 4 The control logic of the proposed VSL control strategy**

     Real-time weather data and traffic data are collected by meteorological stations installed on the roadside and loop detectors on freeway, respectively. These data are used to calculate whether the current taffic risk reaches the start threshold of the VSL control. If the start threshold is reached, the VSL control is started. First, the speed limit of single VSL sign is reduced according to the speed change step, and the target speed limit value of each VSL sign is obtained. Then, the speed limit displayed on VSL sign is revised according to the maximum speed difference between adjacent VSL signs. The speed limit is reduced until the risk is lower than the threshold and then it begins to gradually return to the default value.

*Determination of optimal VSL control factors*
The start threshold of VSL control is set to be 0.2 according to the section of crash risk prediction model under fog conditions. The optimal values of the remaining three control factors are determined by establishing an optimization problem. The goal of optimal VSL control is to effectively reduce traffic safety risks under low visibility without significantly increasing the total travel time. The objective function is determined by:

$$Fitness = -\frac{\Delta R}{\Delta T} \qquad (15)$$

$$\Delta R = \frac{R_{VSL} - R_{Non}}{R_{Non}} \qquad (16)$$

$$\Delta T = \frac{TTT_{VSL} - TTT_{Non}}{TTT_{Non}} \qquad (17)$$

$$R=\sum_{k=1}^{h}\sum_{i=1}^{n}Risk_i(k) \quad (18)$$

$$TTT=\sum_{k=1}^{h}\sum_{i=1}^{n}\frac{l_i q_i(k)}{v_i(k)} \quad (19)$$

where $\Delta R$ and $\Delta T$ are decrease rate of risks and increase rate of total travel time (TTT); $R_{VSL}$ and $R_{Non}$ represent the traffic safety risk with and without VSL control, respectively; $TTT_{VSL}$ and $TTT_{Non}$ denote the TTT with and without VSL control, respectively.

The decision variables are $x = \{T, \Delta v, \Delta V_m\}$. A classic heuristic algorithm, GA, is used to optimize the critical VSL control factors.

## SIMULATION EXPERIMENTS
### Study area
A 2.7-miles section of the I-405 (N) in Orange, California was selected as the study area, as shown in **Figure 5**. It includes three on-ramps and one off-ramp, consisting of one section of six lanes and two sections of five lanes, as shown in **Figure 6**. The inner lane is a High-Occupancy Vehicle (HOV) lane, which was not considered in this study due to its heterogeneity with other lanes.

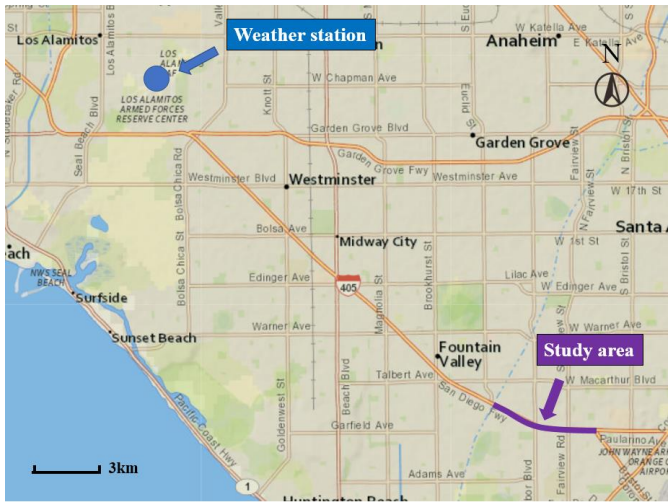

**Figure 5 Study area**

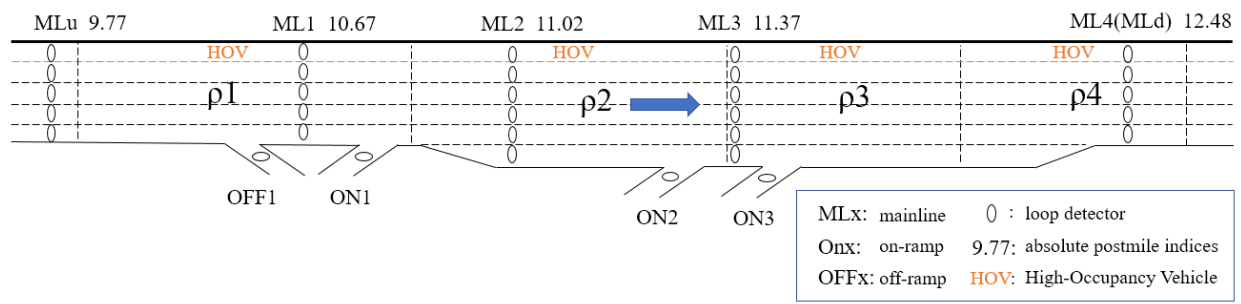

**Figure 6 Schematic diagram of study road section**

### Data preparation

Traffic flow data was collected through underground loop detectors by the freeway Performance Measurement System (PeMS). The detectors record the average speed, traffic volumes and lane occupancy for each lane in each direction every 30 seconds. Data of 7 hours (5:00 am-12:00 am) collected on Tuesday, Thursday, and Saturday from May to July in 2020 was used in the study. Traffic density (veh/mile/ln) of each lane was computed using density = occupancy/g factor, where g factor is the effective vehicle length. For single loop-detector in the study section, PeMS provides a algorithm to calculate g factors(*37*). Then, data of each lane (except HOV lane) were aggregated into a section level.

The road geometric characteristics includes four categories: basic segments (BS), diverging segments (DS), merging segments (MS), and weaving segments (WS). The four road sections in **Figure 6** are respectively DS, MS, MS and BS along the travel direction.

The visibility data was extracted from the National Climate Data Center (NCDC), which provides hourly weather information from weather stations across the United States. The visibility data was extracted from the closest weather station to the study area (**Figure 5**). Then, it was matched with the traffic flow dataset.

Finally, a dataset containing traffic flow data, road geometry data, and visibility data was established.

**Development of crash risk prediction model**

The crash risk prediction model was developed based on the historical data of four freeways (I-5, I-10, I-15, I-405) in California, US during 2013 and 2017, including traffic flow data, fog-related crash data, road geometric data and weather data (*26*). Random forests method was used to rank the importance of variables by Mean Decrease Accuracy (MDA). Variables that significantly affect fog-related accidents were selected, as shown in **TABLE 2**. A Bayesian inference approach based on Markov Chain Monte Carlo (MCMC) method was used to simulate the posterior distribution of $\beta$. The estimation results are shown in **TABLE 3**.

**TABLE 2 Variables Selected for Model Development**

| Category | Symbol | Description |
|---|---|---|
| Weather variable | Visibility | Distance at which object or light can be clearly discerned (mile) |
| Road geometric variable | DS | 1 = if there is a diverge segment; 0 = otherwise |
| Traffic flow variables | AQU | Average vehicle counts at the upstream station (veh/5min) |
| | AQD | Average vehicle counts at the downstream station (veh/5min) |
| | DQU | Standard deviation of vehicle counts at upstream station (veh/5min) |
| | DVU | Standard deviation of speed at upstream station (mph) |
| | DQD | Standard deviation of vehicle counts at downstream station (veh/5min) |
| | DVD | Standard deviation of speed at downstream station (mph) |
| | DV | Speed difference between upstream and downstream stations (mph) |

**TABLE 3 Estimations for the Bayesian Logistic Regression Model**

| Variables | Mean | Std.dev. | 2.5 % | 97.5 % |
|---|---|---|---|---|
| (Intercept) | 0.493 | 1.098 | -1.653 | 2.631 |
| Visibility | -0.202 | 0.067 | -0.337 | -0.073 |
| DS | -1.077 | 0.453 | -1.976 | -0.208 |
| AQU | 0.000 | 0.003 | -0.007 | 0.006 |
| AQD | 0.002 | 0.003 | -0.004 | 0.009 |
| DQU | 0.173 | 0.099 | -0.018 | 0.369 |
| DVU | -0.202 | 0.148 | -0.503 | 0.081 |
| DQD | -0.009 | 0.094 | -0.196 | 0.168 |
| DVD | -0.561 | 0.222 | -1.013 | -0.156 |
| DV | 0.013 | 0.031 | -0.046 | 0.076 |

The mean value of each parameters was selected as the coefficient to calculate the traffic safety risk, as shown in **Equation 20**:

$$Risk_i(k) = 0.493 - 0.202 Visibility - 1.077 DS + 0.002 AQD + 0.173 DQU \\ - 0.202 DVU - 0.009 DQD - 0.561 DVD + 0.013 DV \quad (20)$$

**Development of the MCTM**

The study freeway section was divided into four cells with the same length of 0.8 miles (**Figure 6**). $\Delta T$ was set to 30 s in order that $v_{f,i} \Delta T \leq l_i$ is met. Then, the parameters of each cell were calculated based on the proposed two-step calibration method, including: free-flow speed ($v_f$), backward congestion wave speed ($w_c$), maximum allowable flow ($Q_M$), critical density ($\rho_c$), and jam density ($\rho_j$). **TABLE 4** shows the parameters results. Thus, the MCTM of the section was established.

**TABLE 4 Calibrated Parameters Results of the MCTM**

| Cell | $v_f$ (mph) | $w_c$ (mph) | $Q_m$ (veh) | $\rho_c$ (veh/mile) | $\rho_j$ (veh/mile) |
|---|---|---|---|---|---|
| 1 | 74.25 | 13.55 | 10717.09 | 144.34 | 935.08 |
| 2 | 74.34 | 18.13 | 12478.87 | 167.86 | 856.19 |
| 3 | 85.96 | 14.59 | 11295.81 | 131.41 | 905.48 |
| 4 | 66.39 | 11.06 | 9142.086 | 137.70 | 964.57 |

In order to evaluate the performance of the MCTM, the density of four cells estimated by the MCTM were compared with measured data of the period of 5:00 am to 12:00 am of a day, as shown in **Figure 7**. The mean absolute percentage error (MAPE) and the mean absolute error (MAE) were used to evaluate the prediction performance of MCTM (**Figure 8**). The results show that the MAPEs of four cells are all less than 10%. According to a study by Zhong et al. (*35*), the MCTM developed in this study can be used to simulate the evolution of the traffic flow of the study freeway section. As the underlying traffic simulation model in the VSL control strategy, it can be employed to select the optimal parameters of the VSL control and evaluate the control effects.

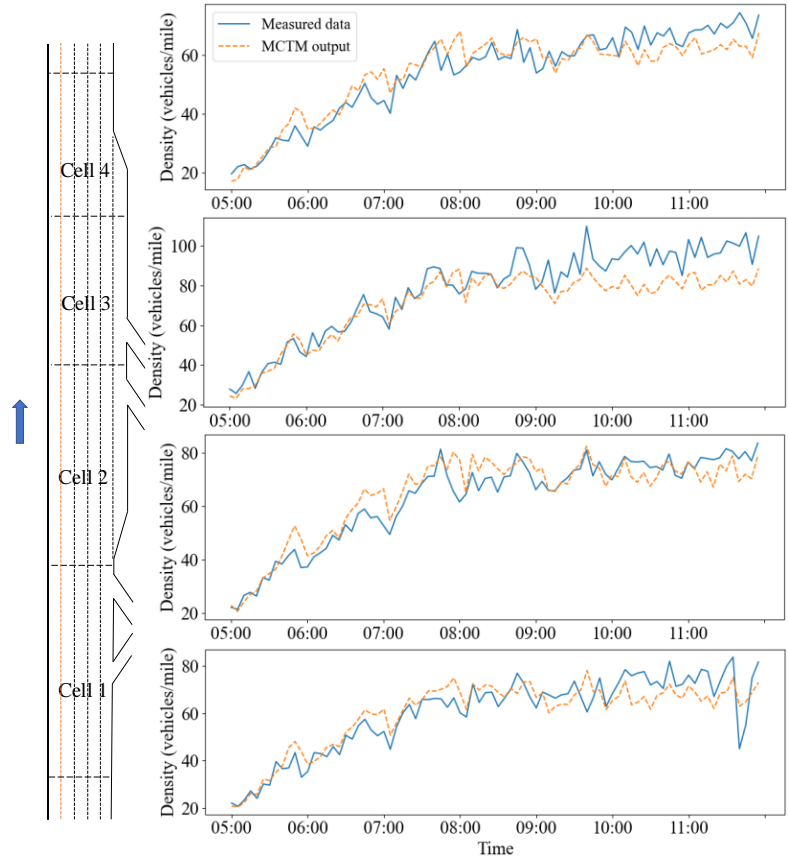

**Figure 7 Density estimation of the MCTM**

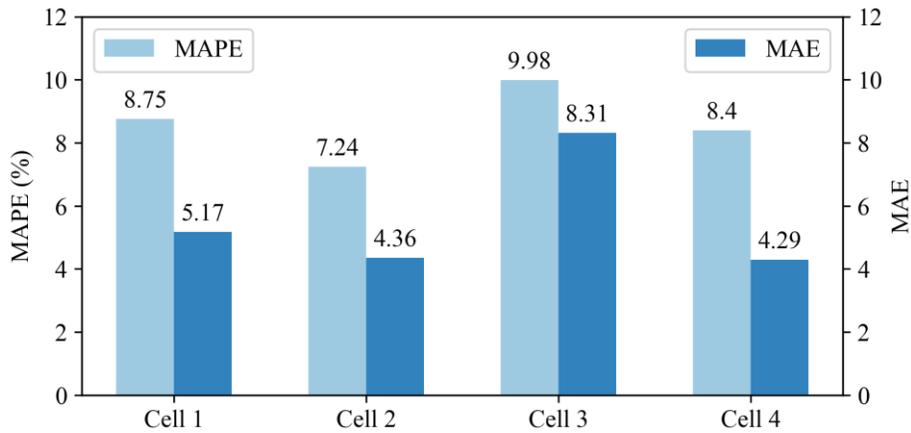

**Figure 8 Prediction performance of the MCTM**

## RESULTS AND DISCUSSIONS
**Optimal VSL control strategy under two weather conditions**
To calculate the optimal VSL control strategy under sunny and fog conditions, the optimization problems for these conditions were established by the proposed method, respectively. According to the weather conditions level, the data of sunny and fog conditions were filtered out. GA in the *scikit-opt* library in

*Ben Zhai, Yanli Wang, Wenxuan Wang and Bing Wu*

*python* was adopted to solve the optimization problems. The key parameters of GA and VSL control are shown in **TABLE 5.**

**TABLE 5 Key Parameters of GA and VSL Control**

| Categoty | Parameters | Value or range |
|---|---|---|
| GA | Objective function dimension | 3 |
|  | Population number | 30 |
|  | Maximum generation | 50 |
|  | Crossover probability | 0.8 |
|  | Mutation probability | 0.1 |
|  | Precision | 1e-7 |
| VSL | Control cycle | 30 seconds to 5 minutes |
|  | Speed change step | 1 to 20 mph |
|  | Maximum speed difference between adjacent signs | 1 to 20 mph |

VSL control strategies with two weather conditions were determined. **TABLE 6** shows the optimal VSL control factors. It can be observed that the optimal start threshold and maximum speed difference between adjacent signs of the two conditions are the same, which are 0.2 and 15 mph, respectively. The optimal control cycle under sunny conditions (150 seconds) is larger than that under fog conditions (120 seconds). And the optimal speed change step under sunny conditions (10 mph) is larger than that under fog conditions (5mph).

**TABLE 6 Optimal VSL Control Factors under Two Conditions**

| Optimal control factors | Description | Conditions | |
|---|---|---|---|
|  |  | Sunny | Fog |
| $R$ (risk value) | Start threshold | 0.2 | 0.2 |
| $T$ (s) | control cycle | 150 | 120 |
| $\Delta v$ (mph) | speed change step | 10 | 5 |
| $\Delta V_m$ (mph) | maximum speed difference between adjacent signs | 15 | 15 |

It can be found that the optimal control cycle and speed change step of the VSL control in fog conditions are lower than those in sunny conditions. It's indicated that the optimal VSL control strategy under fog conditions reduces the speed limit value through a faster change frequency and a smaller speed change step to achieve the safe state. This means that under fog conditions, speed limit should be changed more cautiously.

**Evaluation of VSL control strategy under two weather conditions**
To evaluate the effects of VSL control under two weather conditions, the data of May 15, 2020 and May 18, 2020 were selected as two conditions of sunny and fog, respectively. The effects of the optimal VSL control strategies are shown in **TABLE 7**.

Under sunny conditions, the VSL control significantly reduces the crash risks for cell 3 and cell 4 by 41.32% and 55.74% respectively. In contrast, cell 1 and cell 2 has lower reduction rate of risks, with no more than 2.5%. The total travel time is slightly increased by 2.89%. Under fog conditions, the reduction rate of risks for cell 4 outperforms the other three cells, with reduction by 37.15%. Cell 2 has the lowest risk reduction rate of 5.10%. The VSL control has a lower impact on mobility, only increasing the TTT by 0.48%. It indicates that the VSL control can significantly reduce the risk value of cell 3 and cell 4, compared with no control. The risk changes of cell 4 with and without VSL control is shown as **Figure 9**. It can be seen that two peaks in the risk curve are reduced by VSL control, indicating that the VSL control can reduce high-risk states.

It should be noted that the safety effects of the VSL control under fog conditions are lower than sunny conditions. This is reasonable considering the fact that the optimal speed change step in the VSL control is larger under sunny conditions, which can reduce the risk value more quickly, resulting in a smaller cumulative risk within a certain period. However, due to the large speed change step under sunny conditions, more mobility is sacrificed. Therefore, the increase rate of TTT under sunny conditions is five times greater than that under fog conditions.

**TABLE 7 Effects of the Optimal VSL Control Strategies**

| Conditions | Control effects | Cell 1 | Cell 2 | Cell 3 | Cell 4 | Total |
|---|---|---|---|---|---|---|
| Sunny | Δ% of risk | -2.35 | -2.43 | -41.32 | -55.74 | -25.44 |
|  | Δ% of TTT | 2.64 | 3.25 | 2.83 | 2.86 | 2.89 |
| Fog | Δ% of risk | -10.61 | -5.10 | -14.57 | -37.15 | -16.85 |
|  | Δ% of TTT | 0.11 | 0.04 | 0.26 | 1.55 | 0.48 |

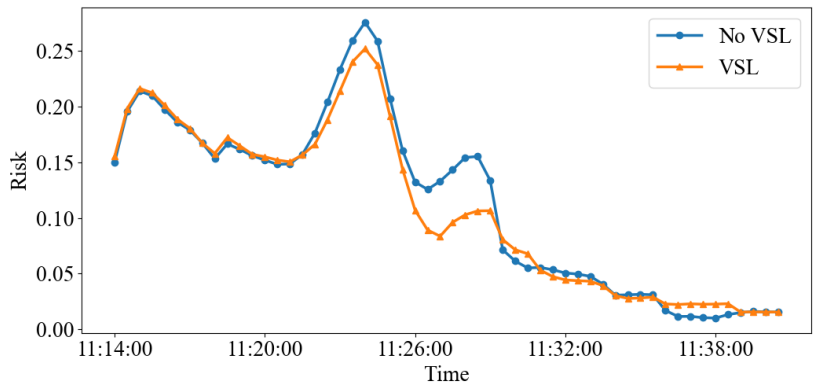

**Figure 9 Risk change of Cell 4 with or without VSL control under fog conditions**

**Effects of placements of VSL signs**
The control effects of VSL are also affected by the placement of VSL signs (*38*). In this section, four placements scenarios of VSL signs were evaluated, as shown of **Figure 10**. In scenario 1, each cell contains a VSL sign, which is the scenario of the control strategy in the above section. Scenarios 2 and 3 both contain two VSL signs, and there are respectively one and two cells between the two signs. Scenario 4 has only one VSL sign, which is set in cell 1. Note that four placements scenarios of VSL signs were evaluated based on their own optimal VSL control strategy under fog conditions.

The results in **Figure 11** suggest that the scenario with larger distance between the two VSL signs brings larger safety improvements. Compared with scenario 1, VSL control in scenario 4 can reduce the risk by about 8%. However, greater safety improvements often mean more mobility sacrifices, as shown in **Figure 11**. This seems reasonable considering the fact that from scenario 1 to 4, the impact of the speed reduction displayed by a single VSL sign is getting wider and wider.

Benefit-cost ratio was used for the comparison between different placements of VSL signs. It is defined as the fitness value of the optimal VSL strategy divided by the number of VSL signs. The results show scenario 1 has the highest benefit-ratio among the four scenarios, and the VSL placement in scenario 1 is recommended on the selected freeway segment.

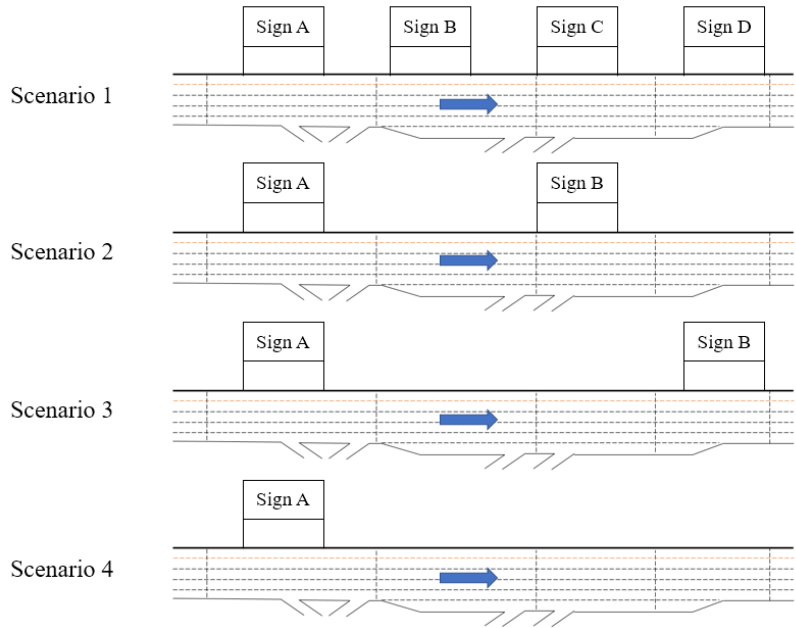

**Figure 10 Different placements of VSL signs**

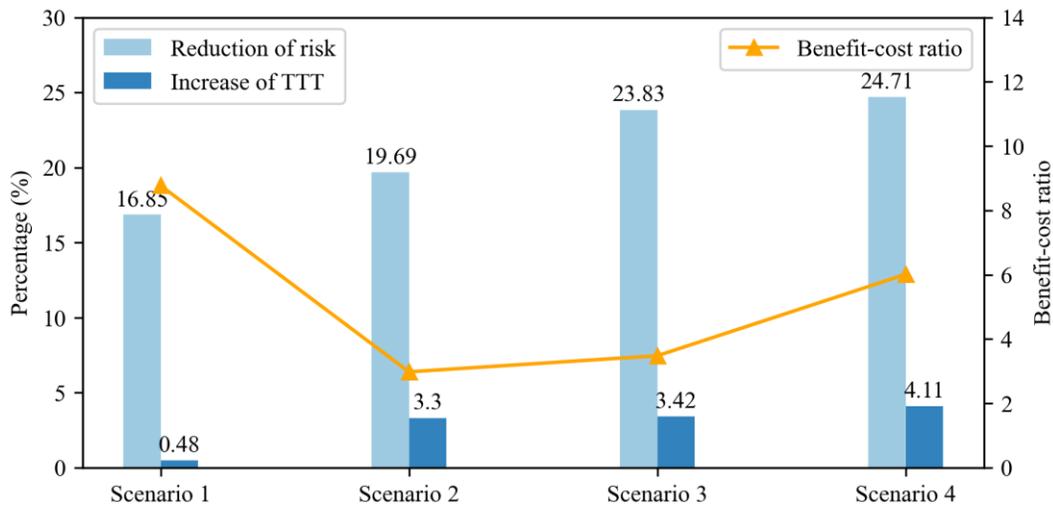

**Figure 11 Effects of the placements of VSL signs**

**CONCLUSIONS**

   Fog has a critical impact on traffic safety on freeway. This study developed optimal VSL control strategies to improve traffic safety on freeway segment under fog conditions. Bayesian logistic regression model was used to develop the crash risk prediction model that established the relationship between crash risks and visibility, traffic flow state, road geometric characteristics. Then, the MCTM was established as the underlying traffic simulation model in the VSL control strategy to simulate the evolution of the traffic flow, and a two-step calibration method of the MCTM was introduced. On this basis, an MPC-based method of determining the optimal VSL control strategy was proposed, which considered safety and mobility. The GA was used to search for the optimal combination of critical factors of the proposed VSL control strategy.

Through a simulation experiments of I-405 (N) in USA, the critical control factors under sunny and fog conditions were determined. In the optimal VSL control strategy under fog conditions, the speed limit changed by 5 mph per 120 seconds. It used 15 mph as the maximum speed difference between adjacent VSL signs. Compared with that under sunny conditions, the optimal VSL control strategy under fog conditions changes speed limit more cautiously. It reduces the speed limit value through a faster change frequency and a smaller speed change step to achieve the safe state.

The control effects under two weather conditions were evaluated. The proposed VSL control strategy effectively improved traffic safety on freeway segment under fog conditions. The optimal VSL control reduced the crash risks by up to 37.15% for a certain cell, and the risk of the entire segment had been reduced by 16.85%. Moreover, it effectively reduced traffic safety risks without significantly increasing the total travel time, with only increasing TTT by 0.48%. Although the optimal VSL control under sunny conditions performed better on crash risk reduction, it resulted in a more increase of TTT compared with fog conditions.

This study also compared the effects of placements of VSL signs. The results show that it significantly affected the control effects of VSL. As the distance between the two VSL signs increased, the crash risks that VSL control can reduce gradually increased, but it also leaded to a decrease in mobility. This is because a longer road section can be affected by a single VSL sign. A benefit-cost analysis indicats that scenario 1 (each cell has one VSL sign) is recommended for field application on the selected freeway segment.

The optimal control strategy of VSL given in this study is expected to be used in active traffic management to help improve traffic safety under fog conditions. However, this study assumed complete drivers' compliances to speed limits posted on VSL signs. In practical applications, the drivers may not obey VSL, and some speed enforcement techniques may need to be combined with VSL control. In addition, to avoid the situation where drivers enter the study road section with a large deceleration, the speed coordination between the first VSL sign and the upstream still needs further study.

## ACKNOWLEDGMENTS

This study is sponsored by the Science and Technology Commission of Shanghai Municipality (STCSM, Grant No.18DZ1200200).

## AUTHOR CONTRIBUTIONS

The authors confirm contribution to the paper as follows: study conception and design: Ben Zhai, Bing Wu; data collection: Ben Zhai; analysis and interpretation of results: Ben Zhai, Yanli Wang; draft manuscript preparation: Ben Zhai, Wenxuan Wang. All authors reviewed the results and approved the final version of the manuscript.